\documentclass[pra,superscriptaddress,twocolumn,showpacs,floatfix]{revtex4}

\usepackage{graphics}
\usepackage{graphicx}
\usepackage{bm}
\usepackage{amsmath}
\usepackage{amsfonts}
\usepackage{amssymb}
\usepackage{latexsym}
\usepackage{color}
\usepackage{soul}
\usepackage{hyperref}

\begin{document}

\title{Repeat-until-success quantum repeaters}

\author{David Edward Bruschi}\thanks{Current affiliation: Racah Institute of Physics and Quantum Information Science Centre, the Hebrew University of Jerusalem, Jerusalem 91904, Israel}
\affiliation{School of Electronic and Electrical Engineering, University of Leeds, Leeds LS2 9JT, United Kingdom}
\author{Thomas M. Barlow}
\affiliation{School of Physics and Astronomy, University of Leeds, Leeds LS2 9JT, United Kingdom}
\author{Mohsen Razavi}
\affiliation{School of Electronic and Electrical Engineering, University of Leeds, Leeds LS2 9JT, United Kingdom}
\email{m.razavi@leeds.ac.uk}
\author{Almut Beige}
\affiliation{School of Physics and Astronomy, University of Leeds, Leeds LS2 9JT, United Kingdom}

\email{m.razavi@leeds.ac.uk}

\date{\today}
 
\begin{abstract}
{We propose a repeat-until-success protocol to improve the performance of probabilistic quantum repeaters. Quantum repeaters rely on passive static linear optics elements and photodetectors to perform Bell-state measurements (BSMs). Conventionally, the success rate of these BSMs cannot exceed 50\%, which is an impediment for entanglement swapping between distant quantum memories. Every time that a BSM fails, entanglement needs to be re-distributed between the corresponding memories in the repeater link. The key ingredient in our scheme is a repeatable BSM. Although it too relies only on linear optics and photo-detection, it ideally allows us to repeat every BSM until it succeeds. This, in principle, can turn a probabilistic quantum repeater into a deterministic one. Under realistic conditions, where our measurement devices are lossy, our repeatable BSMs may also fail. However, we show that by using additional threshold detectors, we can improve the entanglement generation rate between one and two orders of magnitude as compared to the probabilistic repeater systems that rely on conventional BSMs. This improvement is sufficient to make the performance of probabilistic quantum repeaters comparable with some of existing proposals for deterministic quantum repeaters.}
\end{abstract}

\pacs{03.67.Bg, 03.67.Dd, 03.67.Hk, 42.50.Ex}

\maketitle

\section{Introduction} \label{Intro}
{The quest for long-distance quantum communications has resulted in various schemes for quantum repeaters \cite{Zoller_Qrepeater_98, Briegel:repeater_PRA, DLCZ_01, ProbReps:RevModPhys.2011, Razavi.Lutkenhaus.09, Munro:NatPhot:2012, Azuma:All_optical_QR_2013}. Such systems ideally enable two users, at an arbitrarily long distance, to share entangled states, which can consequently be used in applications such as quantum key distribution (QKD) \cite{Ekert_91, BBM_92}, teleportation \cite{BBCJPW_93}, and quantum networking \cite{Kimble:QInternet_Nature}. While there have been successful demonstrations over short distances \cite{Kimble:3mRepeater_Sci07, Pan:300mRepeater_Nat08}, all proposed techniques for quantum repeaters face certain technological challenges for their full implementation. In the originally proposed quantum repeaters \cite{Zoller_Qrepeater_98, Briegel:repeater_PRA}, quantum memories as well as highly efficient deterministic quantum gates are required. These gates enable the Bell-state measurement (BSM) required for entanglement swapping as well as the controlled-NOT operation required for the purification of entangled states \cite{Bennett:purification_PRL, Deutsch:purification_PRL}. In the latest proposals for quantum repeaters \cite{Munro:NatPhot:2012, Azuma:All_optical_QR_2013}, the need for quantum memories, as storage devices, has been eliminated, but high-fidelity quantum processing is still required for the proper operation of such systems. There is also another class of quantum repeaters, known as probabilistic quantum repeaters, which are proposed to alleviate the need for high-fidelity operations by relying on probabilistic gates \cite{DLCZ_01, ProbReps:RevModPhys.2011, Razavi.DLCZ.06, Razavi.Lutkenhaus.09, Razavi.Amirloo.10, LoPiparo:2013}. By restricting the total channel length to a moderate distance up to around 1000~km, probabilistic repeaters are expected to be the first generation of working quantum repeaters, before deterministic or no-memory repeaters come to reality. In this paper, we build on the existing techniques in quantum computing \cite{Beige:RUS_PhysRevLett.2005, Beige:RUS2_PRA2006}, to devise a repeatable technique for BSMs, thereby improving the rate one can achieve for probabilistic quantum repeaters to a level comparable with deterministic ones \cite{vanLoock:repeaterComp_PRA}.   

Probabilistic quantum repeaters rely on quantum memories for entanglement storage, and linear optics and photodetection for entanglement swapping \cite{DLCZ_01, ProbReps:RevModPhys.2011, Razavi.DLCZ.06, Razavi.Lutkenhaus.09, Razavi.Amirloo.10, LoPiparo:2013}. In such systems, the entire channel is split into multiple shorter segments with quantum memories at the two ends of each segment. Entanglement must be initially established over these segments and stored in the corresponding memories. This process is often probabilistic, as it typically relies on single-photon communications \cite{DLCZ_01}, hence requires several repetitions until it succeeds. Once initial entanglement distribution over two neighboring segments is done, entanglement can be extended to farther distances by performing BSMs on the intermittent memories.

In the case of non-interacting memories often used in probabilistic repeaters, a direct BSM on quantum memories is not possible. In this case, one conventionally reads out the memory state, i.e., transfers the memory state to a photonic one, and then performs the BSM over the retrieved photons. The latter measurement can be implemented by using passive static linear optics and photodetection, but, for that matter, is an incomplete measurement with a success rate limited to 50\% \cite{Lutkenhaus:BSMLinear_2001}. This is because that, even if our measurement module is loss free, we can only measure two, out of four, Bell states by such devices. By accounting for the loss in the BSM module, the chance of success would further go down. In the event of a BSM failure, the initial states of the memories cannot be recovered, hence entanglement must be distributed again between the corresponding memories, and the whole process must be repeated. The total rate of entanglement generation may then end up to be low in probabilistic repeaters.

There have been several proposals to improve the BSM success rates on photonic states \cite{Grice_PRA.84.042331, vanLoock_PRL.110.260501}. The main idea is to avoid some of the conditions in the no-go theorem of \cite{Lutkenhaus:BSMLinear_2001}, which enable them to distinguish between more than two Bell states. Auxiliary entangled photons can be used to boost the performance of the BSM (see \cite{Grice_PRA.84.042331}). Under ideal conditions, and using photon-number resolving detectors (PNRDs), it is shown that by the addition of $2^N-2$ entangled photons, the success rate can reach $1-1/2^N$. Conditional feed-forward techniques, relying on a concatenation of linear-optics module, have also been attempted but to no avail \cite{Pavicic_PRL.109.079902}. A more recent proposal used squeezers to make the four Bell states partially distinguishable, which allows for a success rate that exceeds 60\%, under ideal conditions and by using PNRDs. Further proposals rely on nonlinear optics, although these are more challenging to implement \cite{NonlinearBSM_PRL.86.1370}.}

In this work we propose a repeat-until-success (RUS) protocol that can improve the chance of success of the BSMs required for entanglement swapping, therefore improving the performance of probabilistic quantum repeaters. Our protocol relies on passive linear optics and photodetection for performing a BSM. Unlike previously proposed photonic BSMs, our RUS scheme actively incorporates the quantum memories on which the BSM is performed. Instead of simply reading the quantum memory, our scheme uses an entangling procedure between the memory and a photon, known as double encoding \cite{Beige:RUS_PhysRevLett.2005}. The use of entanglement is reminiscent of the technique used in \cite{Grice_PRA.84.042331}, but without requiring external entangled photon sources.
We can then suitably choose measurements of the photon state in such a way that the initial entangled states are not destroyed in case of measurement failure. As we shall see below, using mutually unbiased measurements \cite{Beige:RUS_PhysRevLett.2005}, we can ideally retain the initial entangled states of the memories, if the swap operation is unsuccessful, and repeat the process until success. Under ideal conditions, this makes our probabilistic BSM a deterministic one with unity efficiency, thereby significantly improving the performance of our quantum repeaters.

In the presence of loss, our RUS scheme, too, becomes probabilistic. We can nevertheless achieve higher BSM success rates especially if photon number resolving detectors are available.
Without resolving detectors, it is still possible, by accepting some errors, to perform BSMs, at an improved success rate, by only using typical threshold detectors. By using additional threshold detectors, we would be able to simulate the required photon number resolution property, by which, without loss of the fidelity, the total entanglement generation rate in a probabilistic quantum repeater can be improved by one to two orders of magnitude. That would make the probabilistic approaches to quantum repeaters comparable with those that rely on deterministic, but optimistically realistic, measurement operations \cite{vanLoock:repeaterComp_PRA}.

This work is organised as follows. In Section \ref{theory}, we introduce the basic tools for the practical implementation of the proposed RUS quantum repeater scheme. These include the double-encoding of quantum information in stationary and flying qubits, the implementation of BSMs, and the realization of gate operations between non-interacting quantum memories. Section \ref{ideal} describes a basic RUS quantum repeater link and analyzes its performance under ideal and realistic conditions. In Section \ref{Sec:NRDet}, we introduce two modified RUS schemes, which do not rely on resolving detectors. In Section \ref{more}, we calculate the generation rate of entangled states in a multiple-memory repeater setup using RUS entanglement swapping. Finally, we summarize our findings in Section \ref{conclusions}. 

\section{Basic tools} \label{theory}

In this section, we summarize the building blocks of an RUS quantum repeater. Since our scheme employs the same resources as other proposed probabilistic repeater schemes \cite{DLCZ_01,Razavi.DLCZ.06, Razavi.Amirloo.10, ProbReps:RevModPhys.2011, LoPiparo:2013}, there are many similarities.  Nevertheless, there would be differences in terms of how the state of quantum memories are double-encoded with that of photons, and how the released photons are measured. Moreover, this section introduces a relatively straightforward scheme for the implementation of four different unitary operations, $U_1$ to $U_4$, between two non-interacting quantum memories.

\subsection{Double-encoding of quantum information} \label{sec:DE}

Double-encoding is an entangling process by which the state of a memory is entangled with the state of a single photon. In particular, suppose our memory $A$ is initially in a qubit state $\alpha |0\rangle_A + \beta |1\rangle_A$, where $\alpha$ and $\beta$ are complex numbers that satisfy $|\alpha|^2+|\beta|^2=1$, and $|0\rangle_A$ and $|1\rangle_A$ represent the corresponding basis vectors of the memory. The state of the memory can then be double-encoded by a single photon $a$ according to the following mapping
\begin{equation} \label{double:encoding} 
\alpha |0\rangle_A + \beta |1\rangle_A \rightarrow \alpha |0{\sf V}\rangle_{Aa} + \beta |1{\sf H}\rangle_{Aa},
\end{equation}
where $|{\sf H} \rangle_a$ and $|{\sf V} \rangle_a$ represent orthogonal polarization states of the single photon $a$. As a result of this double-encoding, a photon entangled with the memory would be released, which will be later used for measurement. The above mapping is called double-encoding, as the quantum information content of the initial state has now been encoded into two subsystems.

\begin{figure}[t]
\center
\includegraphics[width=4.5cm]{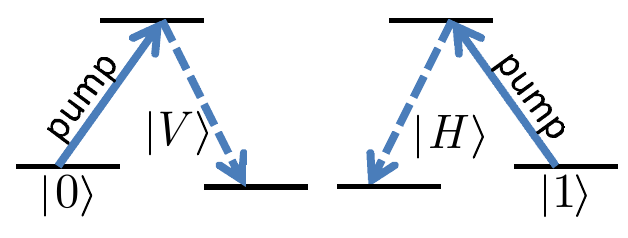} 
\caption{Double-$\Lambda$ atomic structure for the realization of the double encoding scheme described in Eq.~(\ref{double:encoding}). This requires a so-called stimulated Raman adiabatic passage (STIRAP) process, which results in a vertically polarized photon, if the source is initially in $|0 \rangle$ and a horizontally polarized photon, if the source is initially in $|1 \rangle$.} \label{setup}
\end{figure}

One way of implementing the mapping of Eq.~\eqref{double:encoding} is illustrated in Fig.~\ref{setup}. This figure shows an atomic system in the double-$\Lambda$ configuration, where the qubit states $|0\rangle$ and $|1\rangle$ could represent certain hyperfine states. Here we assume that each quantum memory is placed inside an optical cavity, where a stimulated Raman adiabatic passage (STIRAP) for the generation of a single photon on demand can be induced \cite{Kuhn:STIRAP_PRA09}. Two laser pulses are applied such that a qubit initially in $|0 \rangle$ creates a vertically polarized photon and a qubit in $|1 \rangle$ creates a horizontally polarized photon inside the resonator. The photon subsequently leaks out of the cavity and can be used for further processing. Eventually, the atomic state needs to return into its initial state. This can be done, even when the initial state of the quantum memory is not known. 

There are, of course, alternative encoding schemes. For example, depending on the level structure of the respective quantum memories, it might be easier to replace the above polarization encoding with time-bin encoding and to generate photons subsequently. The required BSM will again rely on linear optics elements and photodetectors. Although we focus in the following on polarization encoding, all the operations described below can be implemented analogously with time-bin encoding.

\subsection{An incomplete photonic BSM} \label{sub1}

\begin{figure}[t]
\center
\includegraphics[width=5cm]{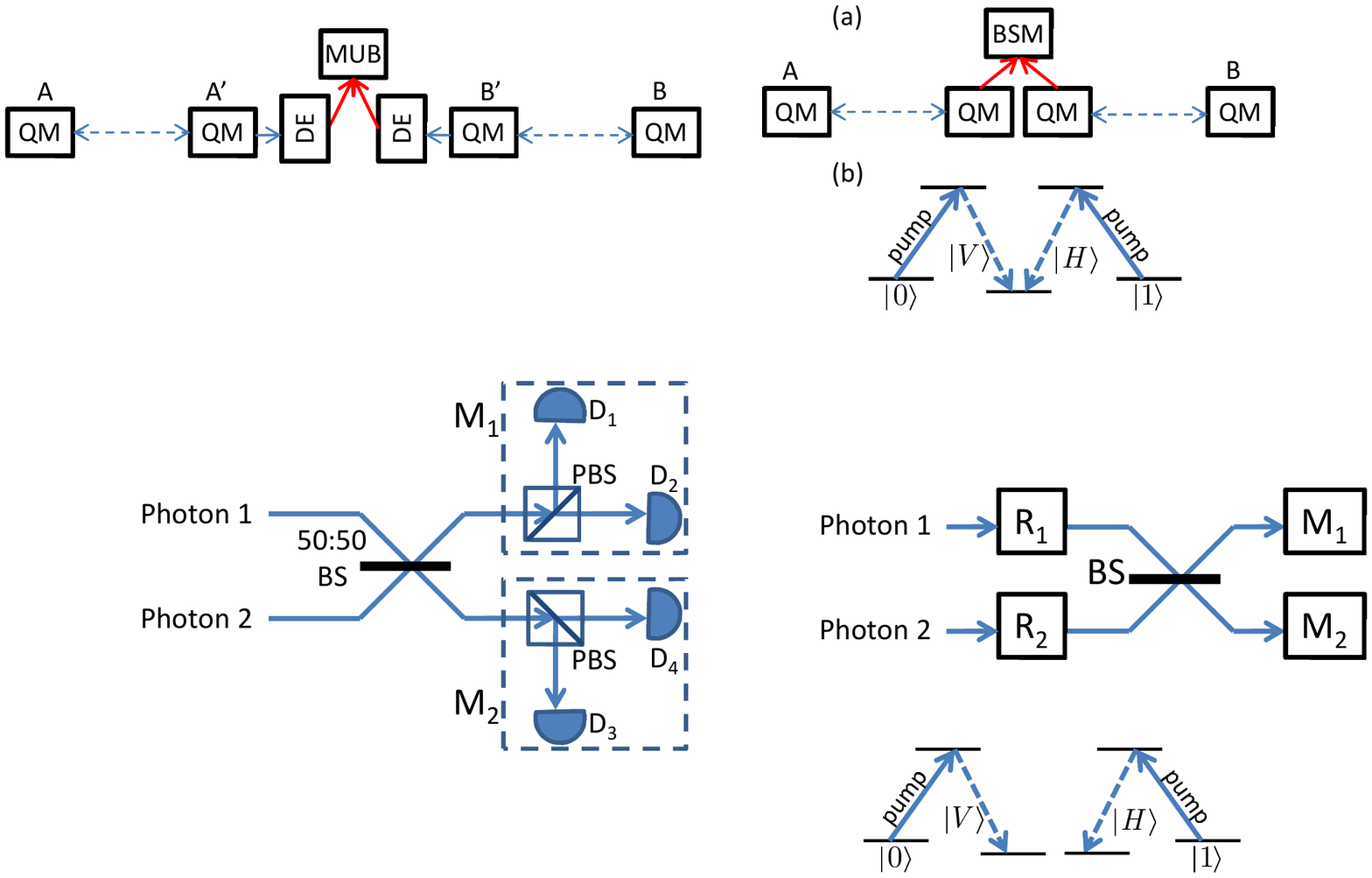} 
\caption{Experimental setup for the implementation of the incomplete photonic BSM described by Eq.~(\ref{basis}) consisting of a 50-50 beamsplitter (BS), two polarising beam splitters (PBS's), and four photodetectors ($D_1$ to $D_4$). In the text, we assume that the PBS's change the path of vertically polarised photons but do not affect horizontally polarised photons.} \label{fig1}
\end{figure}

Passive, static linear optics elements and photodetectors, with infinitely many ancillary vacuum modes, do not allow for the discrimination of four maximally entangled states of two photonic qubits \cite{Lutkenhaus:BSMLinear_2001}. At most, two maximally entangled Bell states and two product states can be distinguished. For example, using the setup of Fig.~\ref{fig1}, it is possible, in the ideal scenario, to distinguish the following photon pair states 
\begin{eqnarray} \label{basis}
&& |\Phi_1 \rangle = |{\sf HH} \rangle \, , ~~ |\Phi_2 \rangle = |{\sf VV} \rangle \, , \nonumber \\
&& |\Phi_{3,4} \rangle = |{\sf HV} \pm {\sf VH} \rangle \, ,
\end{eqnarray}
where, for brevity and throughout this paper, we use a simplified notation which is intuitive and allows us to neglect normalization factors.

In Fig.~\ref{fig1}, the two modes of light enter the measurement module via the two input ports of the beam splitter. The setup is such that photons in either 
$|\Phi_1 \rangle$ or $|\Phi_2 \rangle$ cause clicks at the same detector. More concretely, in the ideal scenario when there is no overall loss, the two input photons, because of quantum interference, would choose the same path at the beam splitter. There will then be clicks on $D_2$ or $D_4$, if $|\Phi_1 \rangle$ is at the input, and clicks on $D_1$ or $D_3$, if $|\Phi_2 \rangle$ is at the input. If we have photon-number resolving detectors, we should be able to count exactly two photons in each case. The two maximally entangled states $|\Phi_3 \rangle$ and $|\Phi_4 \rangle$ are measured when two different detectors click. Clicks on $D_1$ and $D_2$ or on $D_3$ and $D_4$ occur, if the photons are in $|\Phi_3 \rangle$. Clicks on $D_1$ and $D_4$ or on $D_2$ and $D_3$ indicate a measurement of $|\Phi_4 \rangle$, as summarized in Table \ref{table1}. 

\begin{table}[b]
\centering
\begin{tabular}[c]{@{}|c|c|c|}
\hline
Incoming & \multicolumn{2}{|c|}{Measurement outcomes} \\	
photons & Possibility 1 & Possibility 2 \\ \hline 
$|\Phi_1 \rangle $ (or $|\chi_1 \rangle$) & 2 clicks at $D_2$ & 2 clicks at $D_4$ \\ \hline
$|\Phi_2 \rangle $ (or $|\chi_2 \rangle$) & 2 clicks at $D_1$ & 2 clicks at $D_3$ \\ \hline
$|\Phi_3 \rangle $ (or $|\chi_3 \rangle$) & $D_1$ and $D_2$ clicks & $D_3$ and $D_4$ clicks \\ \hline
 $|\Phi_4 \rangle $ (or $|\chi_4 \rangle$) & $D_1$ and $D_4$ clicks & $D_2$ and $D_3$ clicks  \\ \hline
\end{tabular}
  \caption{Overview of the possible measurement outcomes for the photon pair states $|\Phi_i\rangle$, $i=1,\ldots, 4$, considered in Eq.~\eqref{basis} and the photon pair states $|\chi_i\rangle$ considered in Eq.~\eqref{mut}.}
  \label{table1}
\end{table}

\subsection{An incomplete photonic BSM in a mutually unbiased basis} \label{core}

\begin{figure}[t]
\center
\includegraphics[width=5cm]{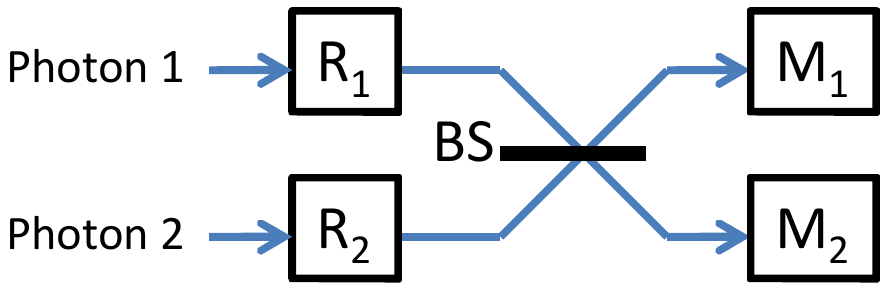} 
\caption{Experimental setup for the measurement of the states given in Eq.~(\ref{mut}), which are the basis vectors of a mutually unbiased basis. Measurement modules $M_1$ and $M_2$ are the same as those in Fig.~\ref{fig1}.} \label{fig2}
\end{figure}

Let us assume that a single qubit rotation is performed before photons 1 and 2 enter the measurement device shown in Fig.~\ref{fig1}. As shown in Fig.~\ref{fig2}, we denote the single-qubit rotation performed on photon $j$ by $R_j$. In the following we assume that 
\begin{eqnarray} \label{Us}
R_j^\dag &=& |{\sf x}_j \rangle \langle {\sf H}| + |{\sf y}_j \rangle \langle {\sf V}| 
\end{eqnarray}
 with $j=1,2$ and with the ${\sf x}$ and ${\sf y}$-states 
\begin{eqnarray} \label{xys}
|{\sf x}_j \rangle &\equiv & |{\sf H} + {\sf V} \rangle \, , ~~ \nonumber \\
|{\sf y}_j \rangle &\equiv & {\rm i}^{j-1}{}|{\sf H} - {\sf V} \rangle  \, .
\end{eqnarray}
Notice that $R_1$ and $R_2$ differ only by the factor ${\rm i}$ in the definition of $|{\sf y}_2 \rangle$. Given that $\langle x_j| y_j \rangle = 0$, the operations in Eq.~(\ref{Us}) are indeed unitary. Both operations, $R_1$ and $R_2$, can be implemented using linear optical elements. 

Proceeding as in the previous subsection, one can show that Fig.~\ref{fig2} now realizes the four projections $|\chi_i \rangle \langle \chi_i|$ with the states $|\chi_i \rangle$ defined as
\begin{eqnarray}
\label{Xis}
|\chi_i \rangle &=& R_1^\dag R_2^\dag \, |\Phi_i \rangle , \quad\mbox{$i=1,\cdots, 4$} \, . 
\end{eqnarray}
Using Eqs.~(\ref{basis})--(\ref{Xis}), we can show that these states are given by
\begin{eqnarray} \label{mut}
|\chi_1 \rangle &=& | {\sf (H+V)(H+V)}  \rangle \, , \nonumber \\
|\chi_2 \rangle &=& {\rm i} \,  | {\sf (H-V)(H-V)} \rangle \, , \nonumber \\
|\chi_3 \rangle &=& | {\sf H(H-{\rm i}V)} + {\sf V({\rm i}H-V)}  \rangle \, , \nonumber \\
|\chi_4 \rangle &=&| {\sf H({\rm i}H-V)} + {\sf V(H-{\rm i}V)} \rangle 
\end{eqnarray}
up to an overall phase factor. This implies that two photons arriving at the same detector now indicate a measurement of the product states $|\chi_1 \rangle$ or $|\chi_2 \rangle$. Analogously, we find that clicks on two different detectors now indicate a measurement of the states $|\chi_3 \rangle$ or $|\chi_4 \rangle$. More details on this measurement scheme are given in Table~\ref{table1}. Similar to $|\Phi_3 \rangle$ and $ |\Phi_4 \rangle$, both states $|\chi_3 \rangle$ and $ |\chi_4 \rangle$ are maximally entangled states. 

\subsection{Unitary operations between quantum memories} \label{operations}

Notice that the photon states in Eq.~(\ref{mut}) are all equal superpositions of the polarisation states $|{\sf HH} \rangle$, $|{\sf HV} \rangle$, $|{\sf VH} \rangle$, and $|{\sf VV} \rangle$. Hence they form a so-called mutually unbiased basis \cite{Wootters:AnnPhys}. Similarly, if one measures the polarization states $|{\sf HH} \rangle$, $|{\sf HV} \rangle$, $|{\sf VH} \rangle$, and $|{\sf VV} \rangle$ in the mutually unbiased basis, the outcome could be any of $|\chi_i \rangle$, $i=1,\ldots,4$ with the same probability. This means that measuring in the $\chi$-basis does not reveal any information about the polarization of the photons. As we shall see below, this plays an important role in our RUS quantum repeater scheme. 

The above measurement, when combined with the double encoding in Section \ref{sec:DE}, can be used to implement unitary operations on quantum memories. Suppose the quantum memories $A'$ and $B'$ are initially in a state of the form
\begin{eqnarray} \label{mutin}
| \psi_{\rm in}\rangle&=& | \alpha \, 00 + \beta \, 01 + \gamma \, 10 + \delta \, 11 \rangle_{A'B'} \, ,
\end{eqnarray}
where the Greek letters denote complex coefficients. The mapping in Eq.~(\ref{double:encoding}) transforms the state of quantum memories and the two generated photons $a$ and $b$ into the state 
\begin{eqnarray}
| \psi_{\rm de}\rangle &=& | \alpha \, 00{\sf VV} + \beta \, 01 {\sf VH} + \gamma \, 10 {\sf HV} + \delta \, 11 {\sf HH} \rangle_{A'B'ab} \, . \nonumber \\
\end{eqnarray}
Finding the photons $a$ and $b$ subsequently in one of the four states $|\chi_i \rangle$ in Eq.~(\ref{mut}) prepares the quantum memories $A'$ and $B'$ in the following state
\begin{eqnarray}
| \psi_{\rm mem} (i) \rangle &=& \langle \chi_i | \psi_{\rm de} \rangle \, . ~~
\end{eqnarray}
More concretely, using the above equations, we find that the states $| \Psi_{\rm mem} (i) \rangle$ can also be written as 
\begin{eqnarray}
| \psi_{\rm mem} (i) \rangle &=& U_i \, | \psi_{\rm in} \rangle \, ,
\end{eqnarray}
where, up to an overall phase factor, the operators $U_i$ are unitary operators given by
\begin{eqnarray} \label{Uis}
U_1 &=& {\rm diag} \, (1,1,1,1) \, , \nonumber \\
U_2 &=& {\rm diag} \, (1,-1,-1,1) \, , \nonumber \\
U_3 &=& {\rm diag} \, (-1,-{\rm i},{\rm i},1) \, , \nonumber \\
U_4 &=& {\rm diag} \, (1,- {\rm i},{\rm i},-1) 
\end{eqnarray}
with respect to the computational basis states $|00 \rangle_{A'B'}$, $|01 \rangle_{A'B'}$, $|10 \rangle_{A'B'}$, and $|11 \rangle_{A'B'}$. While $U_1$ is the identity operator and preserves the initial state of the quantum memory, $U_2$ adds a minus sign when exactly one of the memories is in $|1 \rangle$. In contrast to this, one can show that the quantum gates $U_3$ and $U_4$ are maximally entangling. They rotate maximally entangled states onto product states.

\section{Single-node RUS quantum repeaters} \label{ideal}

One of the fundamental challenges for the realization of commercial QKD is the limited distance over which secure keys can be exchanged \cite{Wang:260kmQKD:2012}. Quantum repeaters are a possible solution to this problem. As mentioned in Sec.~\ref{Intro}, each repeater link consists of a series of nodes. Every intermediary node, in the simplest scenario, is equipped with two quantum memories. We first entangle memories in adjacent nodes. Afterwards, entanglement swapping at the nodes is employed to extend entanglement over larger and larger distances, while destroying the initially generated entangled states between intermediary nodes. The final aim is to create a maximally entangled state shared between the users, which enables them to generate a secret bit of a cryptographic key among other applications. 

\begin{figure}[t]
\center
\includegraphics[width=8cm]{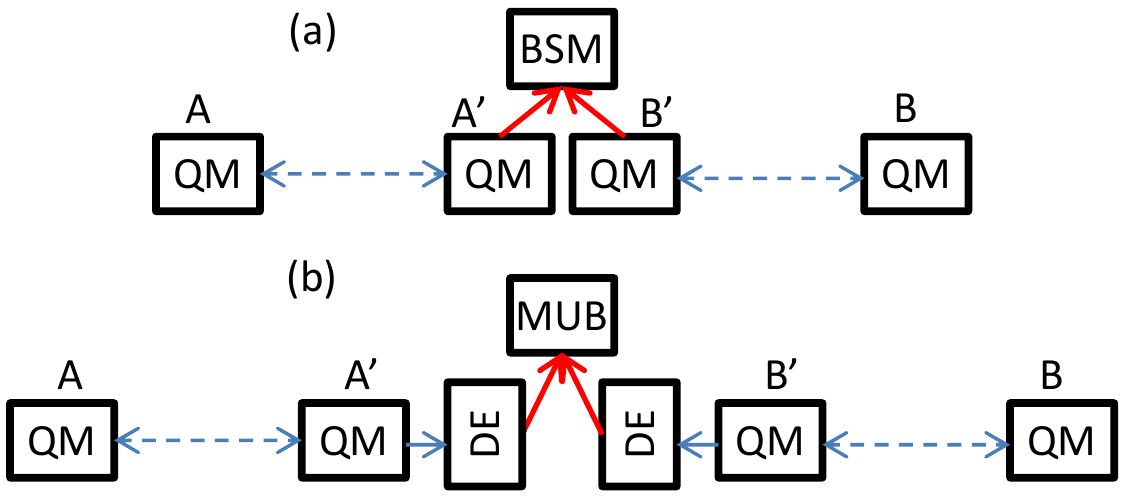} 
\caption{(a) The main building block of a quantum repeater system. Entanglement is first distributed between pairs of quantum memories (QMs) on the left and right sides of the network. A successful BSM on the middle-node QMs can then extend entanglement between remote parties A and B. (b) An RUS quantum repeater link. Instead of directly measuring the retrieved photons from QMs $A'$ and $B'$, we first double encode (DE) their states with photons, and then use the module in Fig.~\ref{fig2} to perform a measurement in the mutually unbiased basis (MUB box).} \label{RUSsetup}
\end{figure}

In this section, we consider the basic building block of quantum repeaters, which is a repeater link with a single middle node. This link is composed of a quantum memory $A$ held by Alice and a quantum memory $B$ held by Bob, respectively, entangled with memories $A^{\prime}$ and $B^{\prime}$ held at the middle node (service provider). In conventional probabilistic quantum repeaters, the BSM on $A^{\prime}$ and $B^{\prime}$ is performed by first reading the states of the memories, that is by transferring the state of each memory to a single photon, and then performing the BSM optically on the released photons by the measurement module in Fig.~\ref{fig1}. Figure~\ref{RUSsetup}(a) shows such a link's setup. In this setup, obtaining states $|\Phi_1 \rangle$ or $|\Phi_2 \rangle$ in Eq.~\eqref{basis} would leave memories $A$ and $B$ in separable states, while the initial entanglement between $A$ and $A'$ as well as that of $B$ and $B'$ are both irreversibly broken. The BSM in Fig.~\ref{RUSsetup}(a) at best has a 50\% chance of success. 

In this section we propose a new {\em repeatable} BSM scheme as shown in Fig.~\ref{RUSsetup}(b). Under ideal conditions, our RUS repeater protocol runs as follows:
\begin{itemize}
	\item[(i)] The provider initializes the pairs of memories $A,A^{\prime}$ and $B,B^{\prime}$ in the maximally entangled state
	\begin{eqnarray} \label{in}
| \Psi_{\rm in}\rangle&=&|01+10\rangle_{AA^{\prime}}\otimes|01+10 \rangle_{BB^{\prime}} 
\end{eqnarray}
using standard techniques \cite{ProbReps:RevModPhys.2011}.
	\item[(ii)] The provider employs the double-encoding scheme, which we described in section \ref{sec:DE}, to entangle memory $A^{\prime}$ with a newly generated photon $a$ and memory $B^{\prime}$ with a newly generated photon $b$. The result of this operation is the following state
\begin{eqnarray} \label{de}
| \Psi_{\rm de}\rangle&=&|0101{\sf HH}+0110{\sf HV} \nonumber\\
&&+1001{\sf VH}+1010{\sf VV} \rangle_{{AA^{\prime}BB^{\prime}ab}} \, .
\end{eqnarray}	
	The memories $A,A^{\prime}$ and photon $a$ now share a tripartite entangled state, so do the memories $B,B^{\prime}$ and photon $b$.
		\item[(iii)] The provider measures the double encoded photons in the mutually unbiased basis. If a suitable measurement outcome is obtained, the four memories $A,A^{\prime},B^{\prime},B$ share a multipartite entangled state. Otherwise, after possibly some local operations, we go back to step (ii). 
	\item[(iv)] The provider separately measures the states of $A^{\prime}$ and $B^{\prime}$ to project the state of the memories $A$ and $B$ into a maximally entangled state. Note that these are single-qubit measurements, which can more easily be done with high precision.
\end{itemize} 
In the following we describe our scheme in more detail and find its performance under ideal and realistic conditions. Under ideal conditions, we show that the success rate in our scheme approaches one if photon number resolving detectors (PNRDs) are used. In the next section, we then propose two practical RUS schemes, which do not rely on PNRDs. 
 
\subsection{RUS entanglement swapping under ideal conditions} \label{ideal:building:block}

In this section, we show how our scheme of Fig.~\ref{RUSsetup}(b) works in the ideal case when there are no inefficiencies in the measurement procedure, all operations are error free, and we are equipped with PNRDs. To model step (iii) in our protocol, we use the results we obtained in Sec.~\ref{operations}. There we showed that the mutually unbiased measurement on photons $a$ and $b$ can be modelled as a unitary operation $U_i$ on quantum memories $A'$ and $B'$. More concretely, using Eq.~(\ref{Uis}), one can show that detection of $|\chi_i \rangle$ in the measurement module of Fig.~\ref{RUSsetup}(b), projects the initial state $|\Psi_{\rm in}\rangle$ in Eq.~(\ref{in}) onto 
\begin{eqnarray}
| \Psi_{\rm mem}(i) \rangle &=& U_i \, |\Psi_{\rm in} \rangle \, \quad\mbox{$i=1,\cdots,4$}, 
\end{eqnarray}
where, using Eq.~(\ref{Uis}), we have 
\begin{eqnarray}
| \Psi_{\rm mem}(1) \rangle &=& |01+10\rangle_{AA^{\prime}}\otimes|01+10 \rangle_{BB^{\prime}} \, , \nonumber \\
| \Psi_{\rm mem}(2) \rangle &=& |01-10\rangle_{AA^{\prime}}\otimes|01-10 \rangle_{BB^{\prime}} \, , \nonumber \\
| \Psi_{\rm mem}(3) \rangle &=& |01 \left(01 + {\rm i} 10 \right) - {\rm i} 10 \left(01 - {\rm i} 10 \right)\rangle_{AA'BB'} \, , \nonumber \\
| \Psi_{\rm mem}(4) \rangle &=& |01 \left(01 - {\rm i} 10 \right) + {\rm i} 10 \left(01 + {\rm i} 10 \right)\rangle_{AA'BB'} \, , \nonumber \\
\end{eqnarray}
up to an overall phase factor. An interesting point in the above equation is that the local unitary operations $U_1$ and $U_2$ do not erase the entanglement between the quantum memories $A$ and $A'$ and between $B$ and $B'$. Finding photons in $|\chi_1 \rangle$ or $|\chi_2 \rangle$ is not a desired measurement outcome. However, the provider no longer needs to return to step (i). He/she only needs to repeat step (ii) until observing either $|\chi_3 \rangle$ or $|\chi_4 \rangle$. This is a great advantage because the initial entanglement distribution part in step (i) is inefficient and costly.

There is a fundamental difference between our approach to BSM enhancement as compared to those proposed in \cite{Grice_PRA.84.042331, vanLoock_PRL.110.260501}. In the latter, the focus is on finding a mechanism by which all four Bell states can be distinguished to some extent. In our approach we are still limited to distinguishing only two Bell states out of four, but our mechanism allows us to repeat the measurement if a product state (which overlaps with one of the two indistinguishable Bell states) is observed. This has become possible by incorporating the quantum memories into our photon pair measurement scheme. Notice that the quantum memories are only indirectly measured, since we use a double-encoding technique.

Measuring the photons in $|\chi_3 \rangle$ or $|\chi_4 \rangle$ is equivalent to the application of a maximally entangling gate to qubits $A'$ and $B'$. In order to complete step (iv) and to disentangle the network node from the quantum memories $A$ and $B$, the provider individually measures the states of the quantum memories $A^{\prime}$ and $B^{\prime}$ in the $|0 \pm 1 \rangle$ basis. For instance, if all four memories are initially in $| \Psi_{\rm mem}(3) \rangle$ and the measurement outcome of the provider is 
\begin{eqnarray} \label{node}
| \Psi_{\rm node}\rangle=|0+1\rangle_{A^{\prime}}\otimes|0+1 \rangle_{B^{\prime}} \, .
\end{eqnarray}
In this case, the memories $A$ and $B$ are left in
\begin{eqnarray}
| \Psi_{\rm final}(3) \rangle&=& |0 \left( 0 + {\rm i} 1 \right) - {\rm i} 1 \left( 0 - {\rm i} 1 \right) \rangle_{AB} \, ,
\end{eqnarray}
which is a maximally entangled state between quantum memories $A$ and $B$. Proceeding analogously, one can show that measuring $A'$ and $B'$ in the $|0 \pm 1 \rangle$ basis, always prepares the remaining quantum memories in a maximally entangled state between the two communicating parties, thereby completing step (iv) of our protocol up to local rotations. This state can now be used to extract one bit of the cryptographic key by performing a local measurement in an appropriately chosen basis. 

\subsection{RUS entanglement swapping under realistic conditions} \label{realistic:building:block}

Under ideal conditions, repeat-until-success entanglement swapping saves time and resources
compared to entanglement swapping with an incomplete BSM. As we have seen above, the memories $A$ and $A^{\prime}$ and the memories $B$ and $B^{\prime}$ need to be entangled only once during the repeater protocol. However, by having a closer look at Table 1, we notice that photon losses would also affect our proposed RUS quantum repeater scheme. As soon as one of the two photons, entangled with memories $A'$ and $B'$, is lost, the measurement outcome becomes ambiguous. For example, a single click at detector $D_1$ might be caused by a photon pair initially prepared in $|\chi_2 \rangle$ or in $|\chi_3 \rangle$. Moreover, without photon number resolving detectors, it is not possible to distinguish two clicks at the same detector from the case, where two photons arrived at different detectors but only one of them caused a click. In this subsection, we have a closer look at the possible imperfections of the measurement module in Fig.~\ref{fig2} and modify step (iii) in our protocol to come up with a basic RUS BSM protocol that works under non-ideal conditions. We calculate the success rate for our basic RUS BSM scheme.  

The main source of imperfection we consider in our analysis is the efficiency of our measurement module. We denote the measurement efficiency of our module by $\eta$. We assume this factor includes the quantum efficiency of our detectors as well as the efficiencies of our double-encoding scheme and any other coupling or path loss in the measurement module. We assume the measurement module is symmetric and we model the measurement efficiency by adding virtual beam splitters with transmissivity $\eta$ before photodetectors. All other elements in the measurement module are then assumed to be loss free. We also introduce a unifying notation for the efficiencies of resolving and non-resolving modules. In our work, $\eta \mu$ denotes the probability of distinguishing two photons, when two photons simultaneously arrive at the same detector. For PNRDs, both photons are equally likely to be detected and $\mu = \eta$. In the case of non-resolving threshold detectors, we have $\mu =0$. In general, $\mu$ can be thought to be between zero and $\eta$. For simplicity, we neglect dark count effects and assume that all single-qubit rotations and measurements can be performed without introducing additional errors.

Step (iii) of our RUS entanglement swapping protocol now has {\em four} possible outcomes. In the first case, two different photodetectors click. Taking into account that every one of the operations $U_i$ occurs with the same probability, one can show that the probability $P_{1+1}$ of this happening, for the initial state as in Eq.~\eqref{in}, is given by
\begin{eqnarray} \label{Pent}
P_{1+1} &=& \frac{1}{2} \eta^2 \, . 
\end{eqnarray}
The second scenario is when two photons arrive at the same detector and the detector registers them both, that is it declares the detection of exactly two photons. The probability $P_{2+0}$ for this to happen is given by
\begin{eqnarray} \label{Pprod}
P_{2+0} &=& \frac{1} {2} \eta \mu \, . 
\end{eqnarray}
Another scenario is the registration of only one click at any of the detectors. This occurs with probability 
\begin{eqnarray} \label{Pprod2}
P_{1+0} &=& P_{1+0} (1) + P_{1+0} (2)
\end{eqnarray}
with $P_{1+0} (1) $ and $P_{1+0} (2) $ given by
\begin{eqnarray} \label{Pprod3}
P_{1+0} (1) &=& \frac{1} {2} \eta (1 - \mu) + \frac{1} {2} (1 - \eta) \eta \ \, , \nonumber \\
P_{1+0} (2) &=& \eta (1 - \eta) \, . 
\end{eqnarray}
Here $P_{1+0} (1)$ accounts for the cases, where the two initial photons are heading to the same detector, while $P_{1+0} (2)$ accounts for the cases where the two photons are heading towards different detectors. In either case, only one detectors has clicked, either because our detectors are non-resolving, or, in the case of PNRDs, because one of the photons has been lost. The final scenario is when no detector clicks. The probability for this to happen is given by
\begin{eqnarray} \label{Pprod4}
P_{0+0} &=& (1 - \eta)^2 \, . 
\end{eqnarray}
One can easily check that the above probabilities add up to one, as they should. Moreover, it is easy to see that the probability of detecting no photon $P_{0+0}$ is much smaller than all other probabilities, if $\eta$ is relatively close to one. For example, for $\eta =0.9$, we have $P_{0+0}$ as small as $0.01$. Later, we will take advantage of this fact.

When including loss and inefficiencies in our model, step (iii) in our protocol has four possible outcomes, where only two of which with certainty specify the output state. In other cases, the post-measurement state can be in a mixture of several states. For instance, suppose, in the setup of Fig.~\ref{RUSsetup}(b), detector $D_2$ clicks (see Fig.~\ref{fig1} for the notation). In this case, we know that the photons $a$ and $b$ are in one of the states $|\chi_1 \rangle$, $|\chi_3 \rangle$, and $|\chi_4 \rangle$ but we cannot say which one. In order to avoid any errors in the final state of the quantum memories $A$ and $B$, in our {\em basic} protocol, we ignore all cases where none, or only a single detector clicks. That is if we register only one photon, or no photon at all, we consider that as a failure, and go back to step (i). The cases where two different detectors click result in a successful entanglement swapping. These cases correspond to the measurement of either $|\chi_3 \rangle$ or $|\chi_4 \rangle$. Moreover, two clicks at the same detector indicate a measurement of $|\chi_1 \rangle$ or $|\chi_2 \rangle$, respectively. In this case, the entanglement between neighboring quantum memories is preserved and the attempted swap operation can be repeated. The total success probability $P_{\rm succ} $ for our basic RUS entanglement swapping scheme hence obeys the relation  
\begin{eqnarray} \label{Psucc}
P_{\rm succ} &=& P_{1+1} + P_{2+0} \cdot P_{\rm succ} \, .
\end{eqnarray}
Taking this and the above probabilities into account, we find that
\begin{eqnarray} \label{Psucc2}
P_{\rm succ}  &=& \frac{P_{1+1}}{1-P_{2+0}}\nonumber\\ &=&\frac{\eta^2}{2 - \eta \mu} \, .
\end{eqnarray}
For PNRDs, and for $\eta > \sqrt{2/3} = 0.816$, the success rate is above 50\%, which beats the maximum that can be achieved by passive static linear optics modules. With existing technology for resolving detectors at efficiencies exceeding 88\% \cite{ResolvingDet_PRA.71.061803}, this is certainly a promising approach for efficient entanglement swapping.

For non-resolving detectors, i.e., for $\mu=0$, the success rate of the above scheme equals
\begin{eqnarray} \label{Psucc3}
P_{\rm succ}  = {1 \over 2} \eta^2 \, . 
\end{eqnarray}
This is exactly the same as the success rate of entanglement swapping schemes based on conventional BSMs as in Fig.~\ref{fig1}. In order to achieve an improvement, the photodetectors need to be able to distinguish one and two photons at least to some extent $(\mu \neq 0)$ or advantage needs to be taken of cases where only one photodetector registers a photon. In the following section, we introduce two protocols that do not rely on PNRDs.

\section{RUS entanglement swapping with non-resolving detectors}
\label{Sec:NRDet}

Our basic RUS entanglement swapping scheme can only beat the 50\% limit if PNRDs are used. Although the technology for superconducting detectors with high efficiency and resolving capabilities is on the rise \cite{ResolvingDet_PRA.71.061803}, it would be more practical if our RUS scheme could work with threshold non-resolving single-photon detectors as well. In this section, we propose two modified RUS BSM schemes, which do not rely on PNRDs. In the first approach, we use the same setup as in Fig.~\ref{fig2} with non-resolving detectors, but allow for errors in the final state. In our second scheme, we create resolving capability by concatenated splitting followed by an array of threshold detectors.

\subsection{Modified RUS BSM with possible errors}

One way of improving the success rate in our basic protocol is to not abort the protocol if only one detector clicks. Note that in this section we assume all detectors are non-resolving, i.e., $\mu = 0$. In this case, a single click might be because of two photons heading toward the same detector with a click probability $P_{1+0}(1)$, or two photons heading toward different detectors with a click probability $P_{1+0}(2)$. For values of $\eta$ near 1, the former case is much more likely than the latter. That is because, after including loss, in the former, both one-photon and two-photon components can make a click, whereas in the latter only one-photon terms can generate only one click. Therefore by accepting some errors, in step (iii) of our basic protocol, we can interpret the detection of only one photon as a measurement of either $|\chi_1 \rangle$ or $|\chi_2 \rangle$. Instead of aborting the protocol, when only one of the photodetectors clicks, we now continue until either none or two detectors click. We now only abort the entanglement swapping, if no detector clicks, which is relatively unlikely to occur (cf.~Eq.~(\ref{Pprod4})).
 
A closer look at Table 1 shows, for example, that a click on the non-resolving detector $D_2$ prepares the quantum memories in the following statistical mixture
\begin{eqnarray} \label{perror1}
\rho_{\rm mem}(1) &=& {1 \over P_{1+0}} \, \nonumber \\
&& \times \big[ \left( P_{1+0}(1) \right) \, |\Psi_{\rm mem}(1) \rangle \langle \Psi_{\rm mem}(1) | \nonumber \\
&& + {1 \over 2} P_{1+0}(2) \, |\Psi_{\rm mem}(3) \rangle \langle \Psi_{\rm mem}(3) | \nonumber \\
&& + {1 \over 2} P_{1+0}(2) \, |\Psi_{\rm mem}(4) \rangle \langle \Psi_{\rm mem}(4) | \, \big] \, . ~~~
\end{eqnarray}
Alternatively, this density matrix can be written as 
\begin{eqnarray} \label{perror2}
\rho_{\rm mem}(1) &=& (1 -  P_{\rm error} ) |\Psi_{\rm mem}(1) \rangle \langle \Psi_{\rm mem}(1) | \nonumber \\
&& + {1 \over 2} P_{\rm error} \, |\Psi_{\rm mem}(3) \rangle \langle \Psi_{\rm mem}(3) | \nonumber \\
&& + {1 \over 2} P_{\rm error} \, |\Psi_{\rm mem}(4) \rangle \langle \Psi_{\rm mem}(4) | \, ,
\end{eqnarray}
where $P_{\rm error} = P_{1+0}(2)/P_{1+0}$ denotes the probability of not preparing $|\Psi_{\rm mem}(1) \rangle$, although we assume that this is the case. Comparing Eqs.~(\ref{perror1}) and (\ref{perror2}) and combining this result with Eqs.~(\ref{Pprod})--(\ref{Pprod3}) yields
\begin{eqnarray}
P_{\rm error} &=& {2(1- \eta) \over 4 - 3 \eta}. 
\end{eqnarray}
When $\eta$ tends to one, $P_{\rm error}$ tends to zero. This is not surprising, since, at $\eta = 1$, a single click can only be caused by the two photons heading toward the same detector.

Proceeding analogously, one can check that the same density matrix $\rho_{\rm mem}(1)$ is obtained in the case of a click at detector $D_4$. Moreover, one can show that the same error rate applies with respect to the preparation of $|\Psi_{\rm mem}(2) \rangle$. The preparation of an analogous density matrix $\rho_{\rm mem}(2)$ with the same error rate occurs in the case of a click at detectors $D_1$ or $D_3$. Within some approximations, the success rate for our modified RUS BSM scheme, i.e., the probability to eventually get two clicks on two different detectors, then obeys
\begin{eqnarray} \label{Psuccx}
P_{\rm succ} &=& P_{1+1} + \left( P_{1+0} \right) P_{\rm succ} 
\end{eqnarray}
which yields
\begin{eqnarray} \label{Psucc3x}
P_{\rm succ}  &=& {P_{1+1} \over 1- P_{1+0}} =  \frac{\eta}{4 - 3 \eta} . 
\end{eqnarray}
For example, for $\eta = 0.9$, there is now roughly a 69.2\% success rate for establishing entanglement between $A$ and $B$, comparable to the over 68\% success rate for our basic protocol when PNRDs are used. At $\eta = 0.9$, the success probability for our basic protocol when threshold detectors are used is only 40.5\%, which is the same as conventional incomplete BSMs. This is not surprising, since there is only a small probability for not seeing any clicks, in which case we abort our modified protocol.

Finally, let us have a closer look at the price that we have to pay for the improvement in the efficiency of our modified protocol. Here, we obtain a lower bound for the fidelity of the final state of the quantum memories shared by Alice and Bob by making the worst-case assumption. That is, in the case of single clicks, we assume that the fidelity obtained from the error terms in Eq.~\eqref{perror2} is zero. In this case, the final fidelity $F$ obeys
\begin{eqnarray} \label{fid}
F &=& P_{1+1} + P_{1+0}(1) ~F, \end{eqnarray}
which results in
\begin{eqnarray} 
F &=& \frac{\eta}{2-\eta}.  
\end{eqnarray}
For example, for $\eta = 0.95$, the communicating parties obtain the desired maximally entangled state with a fidelity exceeding $90\%$. At $\eta = 0.9$, we have a fidelity $F$ over $81.8\%$. This is still above the $2/3$ fidelity that we need for applications like teleportation, although it may not be sufficient for applications in QKD. In the latter case, the quantum bit error rate is roughly given by $P_{\rm error} /2$, which is, respectively, about 8\% and 4\% for $\eta = 0.9$ and $\eta = 0.95$. Such an error and the required distillation that it requires often washes away the additional gain we can obtain from this modified RUS scheme. In the next subsection, we propose another modified RUS scheme for threshold detectors without generating any fundamental errors.

\subsection{Modified RUS BSM with no errors} \label{bad}

\begin{figure}[h!]
\center
\includegraphics[width=\linewidth]{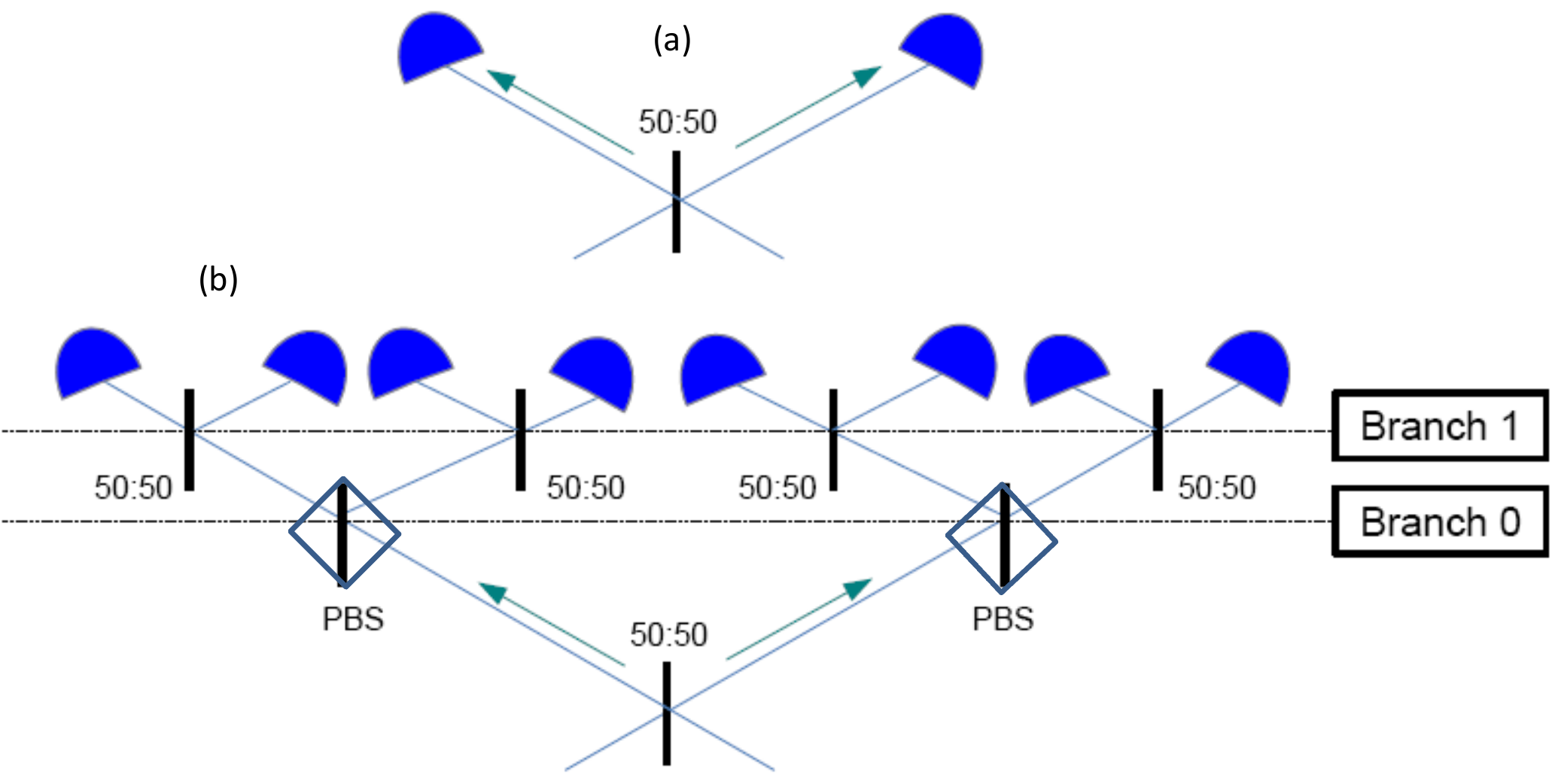} 
\caption{(a) The main branching module. At each branching level, each detector in the previous branch is replaced by this module. (b) The first level of branching when every detector in Fig.~\ref{fig1} has been replaced by the module in (a).} \label{Nesting}
\end{figure}

In this subsection, we discuss another way of improving the success probability of RUS quantum repeaters with threshold detectors without sacrificing the fidelity of the distributed entangled state. Here, some photon-number resolving capability, as required in our basic protocol, is achieved by using concatenated splitting followed by an array of non-resolving single-photon detectors. In this modified scheme, we follow the same procedure as in the basic protocol in Sec.~\ref{basic}, but we replace each detector in Fig.~\ref{fig2} with the module in Fig.~\ref{Nesting}(a). By doing so, there is a 1/2 chance that a two-photon state would split at the 50-50 beam splitter of Fig.~\ref{Nesting}(a), hence creating a double click, which will be similar to registering two photons on a resolving detector. In fact, this new module can be thought as a resolving detector with $\mu = \eta/2$. The total repetition probability, for the resulting module in Fig.~\ref{Nesting}(b), will then be given by $P_{2+0} = \frac{1} {4} \eta^2$, which is now nonzero. We can use the same idea again, by replacing each detector in Fig.~\ref{Nesting}(b) with the module in Fig.~\ref{Nesting}(a), to further improve the repeat probability. If we repeat this branching technique $N$ times, the probability that both photons after $N$ rounds of splitting still impinge on the same detector is given by $\frac{1}{2^{N}}$. For $N$ levels of branching, we then obtain  
\begin{eqnarray} \label{branching:probability}
P_{2+0} &=& \frac{\eta^2}{2}(1-\frac{1}{2^{N}}),
\end{eqnarray}
which reduces to Eq.~\eqref{Pprod} at $\eta = \mu$ for PNRDs when $N \rightarrow \infty$. This is consistent with the well known result that one can achieve a perfectly resolving detector by employing a very large number of non-resolving detectors. Notice that Eq.~\eqref{branching:probability} also applies to the scenario when there is no branching ($N=0$). In this case, it is impossible to use non-resolving detectors to distinguish between two photons arriving via the same branch. 

\begin{figure}[h!]
\center
\includegraphics[width=7.6cm]{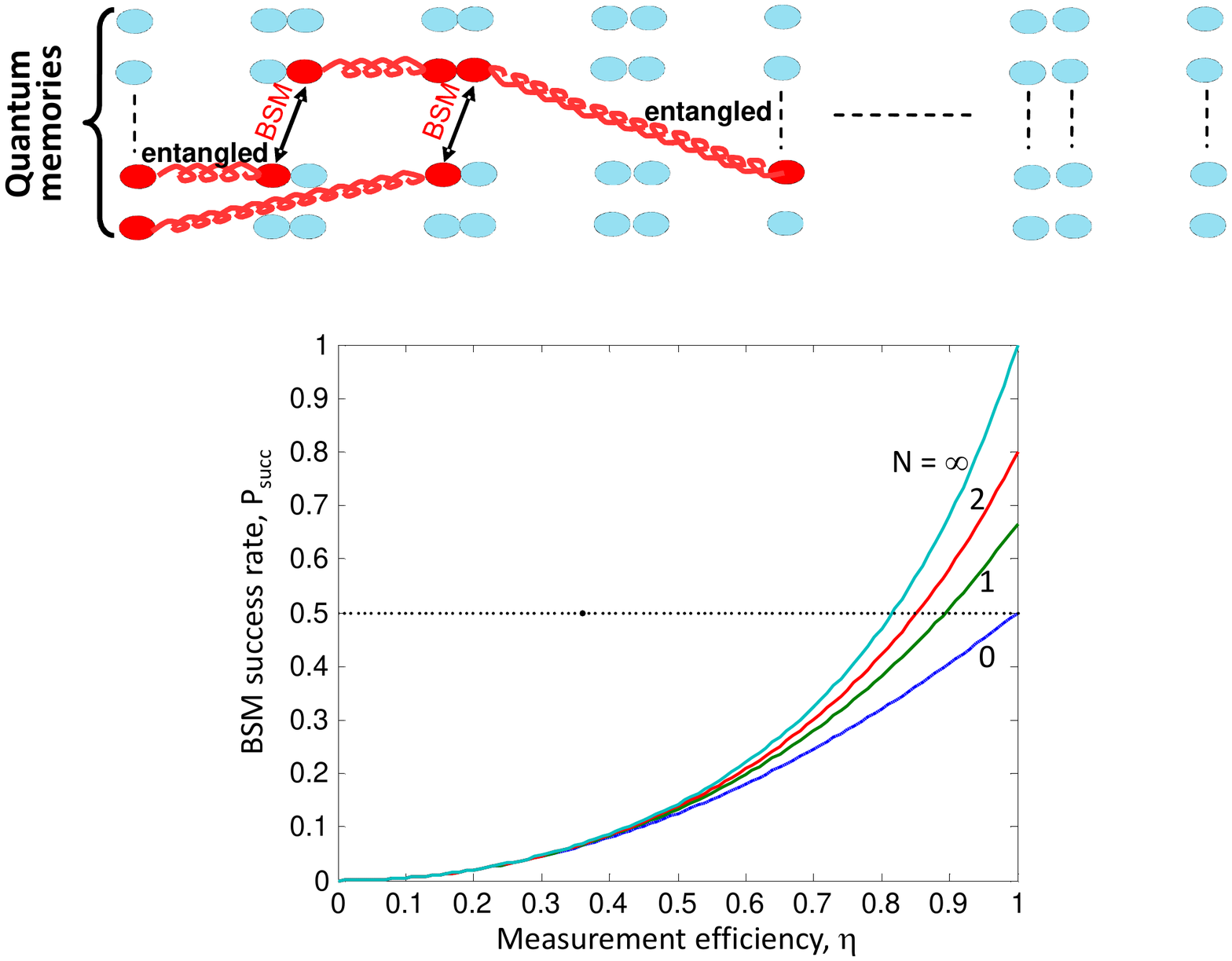} 
\caption{Success probability for the RUS BSM versus the detector efficiency for different branching levels $N$. The curve for $N=0$ represents the case of a non-RUS scheme, whereas $N=\infty$ corresponds to the case when resolving detectors are used.} \label{Fig:Psucc}
\end{figure}

Figure \ref{Fig:Psucc} shows the success probability for our RUS BSM for different branching levels. The curve for $N=0$ represents the no-branching case, where $P_{\rm succ} = \eta^2/2$ takes a maximum value of 1/2. For all other curves, we can surpass this limit for sufficiently large values of $\eta$. For instance, at $N=2$, for $\eta$ greater than 0.853, we are in the region that is not achievable by the BSM module of Fig.~\ref{fig1}. Note that with today's technology we can use single-photon detectors with efficiencies in excess of 90\% \cite{Nam_NatPhot_93p_2013}. The curve labeled $N=\infty$ corresponds to the case where either resolving detectors are employed, or non-resolving detectors are used after infinitely many levels of branching. Practically, with 4 to 5 levels of branching, we can almost achieve the same performance that can be achieved by resolving detectors. 

Several points are worth mentioning regarding the practicality and the cost of our detector-array solution. At first glance, this approach may seem to some extent impractical or costly. One should, however, note that our BSM measurement is a local one, which implies that all detectors needed for the measurement and their corresponding electronics can, in principle, be fabricated on the same platform, and many peripherals of such a system may be shared among all detectors to cut costs. Furthermore, one can think of a simplified structure in which the detector array is replaced by a series of delay lines and a fast optical switch followed by only one non-resolving single-photon detector. Knowing that there are only two photons, in total, to be detected, the delay lines can separate them in time. If we use fast detectors with low deadtime values \cite{GHz_QKD_09}, we then just need one detector and a switch to guide all the photons sequentially toward that detector. If the pulse width of the input photons is on the order of nanoseconds or shorter, the additional loss due to delay lines should be negligible, and our above analysis would hold to some good approximation. In short, we can benefit from several implementation tricks to reduce the cost of implementation for our modified scheme.

\section{Many-node RUS quantum repeaters} \label{more}

We can apply the RUS entanglement swapping protocol described in Sec.~\ref{bad} to a full quantum repeater setup. The improved BSM success probability would in particular help probabilistic structures for quantum repeaters \cite{DLCZ_01,Razavi.Amirloo.10, ProbReps:RevModPhys.2011, LoPiparo:2013}, which find applications in QKD. Here, we consider the multiple-memory quantum repeater structure described in \cite{Razavi.Lutkenhaus.09} to make a comparison between RUS and non-RUS quantum repeater protocols. 

\begin{figure}[h!]
\center
\includegraphics[width=8.6cm]{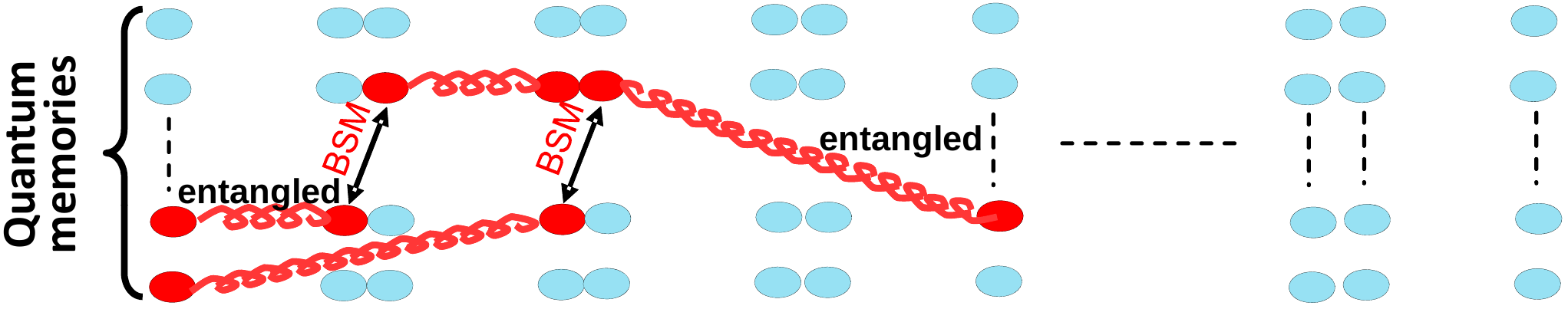} 
\caption{A multiple-memory quantum repeater. In each round entanglement distribution is attempted between any pairs of memories in adjacent stations that are not yet entangled. Once relevant middle stations are informed of entanglement distribution over shorter segments, they apply entanglement swapping on the corresponding memories to further extend the entanglement. In our RUS quantum repeater, this stage is being performed by using RUS BSM operation.} \label{Qrepnet}
\end{figure}

Figure \ref{Qrepnet} shows the structure of a multiple-memory quantum repeater network. In this setup, the total distance $L$ has been split into shorter segments of length $L_0$. We assume that there are a large number of memories at the two ends of each segment. In each round, an entanglement distribution scheme is employed to entangle pairs of memories in neighboring nodes that are not yet entangled. This is typically a probabilistic process, which we denote its success probability by $P_S$. This parameter is commonly proportional to the loss in the channel \cite{Rempe:Nature:2012}, and, for our numerical calculations, we assume $P_S = 0.1 \exp(-L_0/L_{\rm att})$, where $L_{\rm att}$ is the attenuation length of the channel, and the prefactor 0.1 accounts for other possible imperfections in the process. Each entangling attempt typically requires a minimum time of $T_0 = L_0/c$, where $c$ is the speed of light in the channel. We employ the cyclic protocol in \cite{Razavi.Lutkenhaus.09}, where, in each $T_0$-long round, in addition to entangling memories in neighboring nodes, we also extend entanglement by performing BSMs on relevant memories in the middle nodes. The key requirement here is that these BSMs must be {\em informed}, that is, we should know whether the corresponding memories are entangled with some other memories in other nodes. In this case, in the limit of a large number of memories, the rate at which we can entangle memories over a distance $L = 2^n L_0$ per logical memory used is given by \cite{Razavi.Lutkenhaus.09}
\begin{equation}
R_{\rm ent} = \frac{P_S (P_{\rm succ})^n}{2L/c},
\end{equation}  
where $n$ is the nesting level of the repeater setup and $P_{\rm succ}$ is the BSM success probability, here, assumed to be the same for all nesting levels. In our case, $P_{\rm succ}$ can be obtained from Eqs.~\eqref{Psucc2} and \eqref{branching:probability}.

\begin{figure}[ht!]
\center
\includegraphics[width=7.6cm]{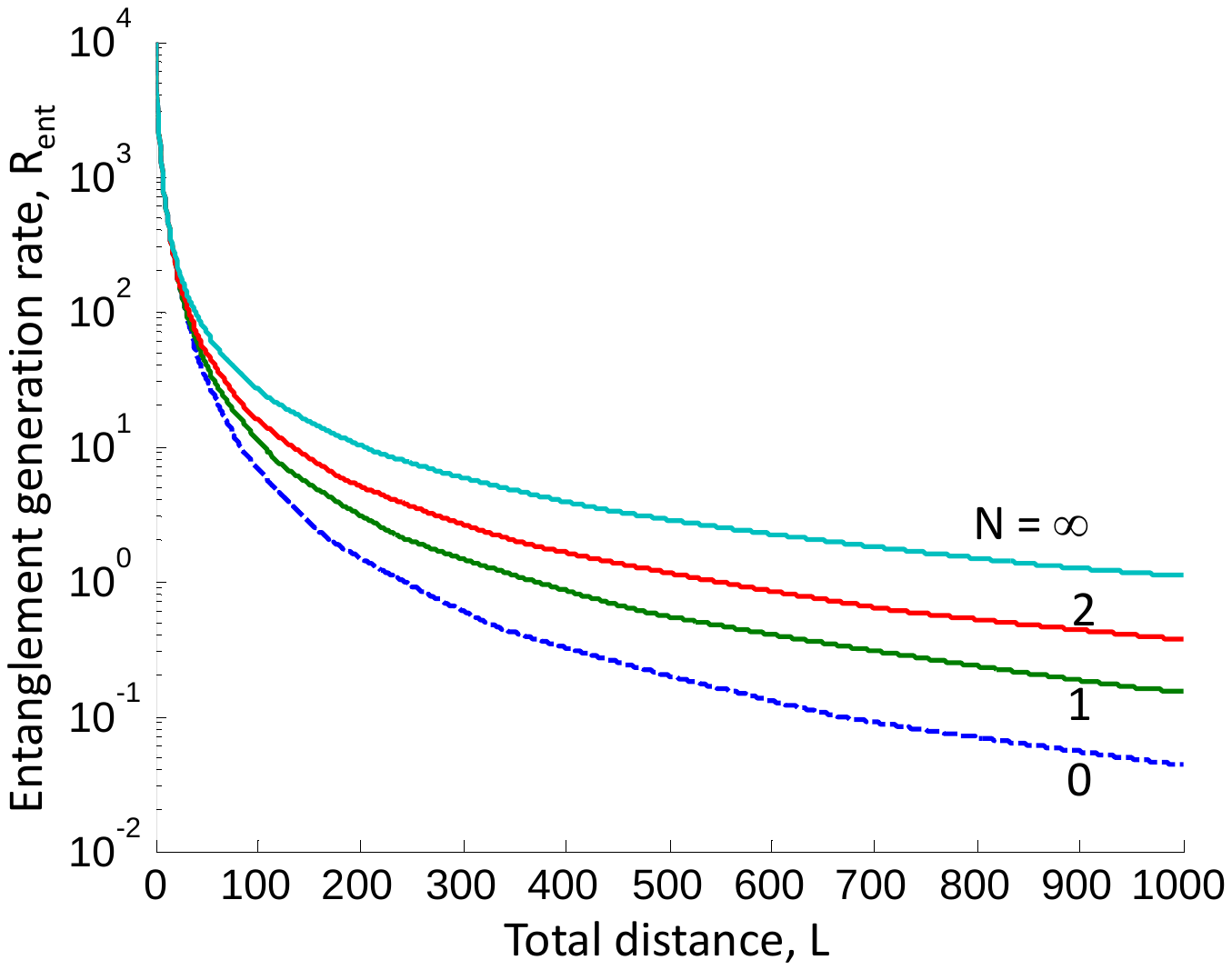} 
\caption{Entanglement distribution rate (1/s) over a distance $L$ for different branching levels $N$. The dashed line represent the case of a non-RUS scheme, whereas $N=\infty$ corresponds to the case when resolving detectors are used. In all curves, and, at each point, the optimum value of $n$ is used. We assume $\eta = 0.93$, $c = 2 \times 10^5$~km/s, and $P_S = 0.1\exp(-L_0/L_{\rm att})$, where $L_{\rm att} = 25$~km.} \label{Fig:Rent}
\end{figure}

Figure \ref{Fig:Rent} shows the number of entangled pairs per memory per second that can be generated for the quantum repeater network of Fig.~\ref{Qrepnet} at a distance $L$. We have considered the non-RUS scheme (dashed line) that uses the BSM module in Fig.~\ref{fig1} for its entanglement swapping, and those of RUS schemes with different levels of branching. In each curve, we have plotted the rate at the optimum value of the nesting level $n$. Given that the RUS entanglement swapping is more efficient than the non-RUS one, the optimum nesting level for the former is typically higher than that of the latter. For instance, at $L = 1000$~km, the optimum value of $n$ for the non-RUS scheme is 5, whereas for the RUS scheme, with $N=1,2$, it is 6, and eventually 7 as $N \rightarrow \infty$. From Fig.~\ref{Fig:Rent}, it can be seen that, by using our proposed RUS scheme for entanglement swapping, we can achieve one to two orders of magnitude improvement in the entanglement generation rate per memory used. This would translate in either requiring proportionally fewer number of memories, or achieving higher secret generation rates once this setup is used for QKD applications.
 
\section{Conclusions} \label{conclusions} 

Entanglement swapping is a key operation for quantum repeater networks. Most conventional approaches to entanglement swapping rely on passive static linear-optics modules, which cannot perform a full BSM. That is, in over 50\% of the time, an inconclusive result is obtained, which requires the repetition of time consuming initialization steps in quantum repeaters. In our work, we used the repeat-until-success protocol to improve the success rate for such modules once used in a repeater setup. Quantum repeaters rely on quantum memories and entanglement swapping between these memories. Our protocol relied on an entangling procedure between memory states and photons followed by an optical measurement in mutually unbiased bases. The combination of these two allowed us to repeat the required BSM in some cases of failure, thereby achieving higher than 50\% success rates. The success rate will approach one in the ideal case and when resolving photodetectors are used. Photon number resolution could be achieved by sequential splitting of photons and detection with an array of non-resolving detectors. In the latter case, a few stages of splitting was sufficient to improve the entanglement generation rate by over one order of magnitude at a nominal distance of 1000~km as compared to non-RUS schemes.

\section*{Acknowledgments}
DEB and MR acknowledge financial support in part by the UK Engineering and Physical Sciences Research Council Grant No.~EP/J005762/1 and the European Community's Seventh Framework Programme under Grant Agreement 277110. TMB receives funding from a UK White Rose Studentship Network on Optimising Quantum Processes and Quantum Devices for future Digital Economy. 

\bibliographystyle{apsrev}
\bibliography{Bibli28Sept12}
\end{document}